\documentstyle[epsfig]{aipproc}
\newcommand{\MeV}{{\rm\,MeV}}
\newcommand{\GeV}{{\rm\,GeV}}
\newcommand{\slk}{/\kern-6pt k}
\newcommand{\slp}{p\kern-5pt/}

\newcommand{\pfrac}[2]{\left(\frac{#1}{#2}\right)}
\begin{document}
\title{On $t\bar t$ threshold and\\[3pt] top quark mass definition}
\author{Oleg Yakovlev$^+$ and Stefan Groote$^*$}
\address{$^+$Randall Laboratory of Physics, University of Michigan,
Ann Arbor, Michigan 48109-1120, USA\\
$^*$Institut f\"ur Physik der Joh.-Gutenberg-Universit\"at,
  Staudinger Weg 7, 55099 Mainz, Germany}
\maketitle
\begin{abstract}
In this talk we discuss the process $e^+e^-\to t\bar t$ near threshold,
calculated with NNLO accuracy as well as top quark mass definitions adequate
for the threshold. As new definition we suggest the $\overline{\rm PS}$ mass
which allows to calculate recoil corrections to the static PS mass. Using this
result we calculate the cross section of $e^+e^-\to t\bar t$ near threshold at
NNLO accuracy adopting three alternative approaches, namely (1) fixing the pole
mass, (2) fixing the PS mass, and (3) fixing the new mass which we call the
$\overline{\rm PS}$ mass. We demonstrate that perturbative predictions for the
cross section become much more stable if we use the PS or the
$\overline{\rm PS}$ mass for the calculations. A careful analysis suggests
that the top quark mass can be extracted from a threshold scan at NLC with an
accuracy of about $100 \MeV$.
\end{abstract}\vfill

\section{Introduction}

In this talk we discuss two topics. We demonstrate how to calculate the cross
section for process $e^+e^-\to t\bar t$ near threshold for NNLO accuracy and
discuss a top quark mass definition adequate for the threshold. 

Top quark physics will be one of the main subjects of studies at future
$e^+e^-$ and $\mu^+\mu^-$ colliders such as the Next Linear Collider
(NLC)~\cite{Peskin} and the Future Muon Collider (FMC). The goals are to
measure and to determine the properties of the top quark which was first
discovered at the Tevatron~\cite{Fermilab} with a mass of
$m=174.3\pm 5\GeV$~\cite{Broojmans}. Although the top quark will be studied at
the LHC and the Tevatron (RUN-II) with an expected accuracy for the mass of
$3\GeV$, the {\em most accurate measurement\/} of the mass with an accuracy
of $0.1\%$ ($100 \MeV$) is expected to be obtained only at the
NLC ~\cite{Review,YakovlevGroote}.
  
The top quark mass enters the relation between the electroweak precision
observables indirectly through loops effects. The global electro-weak fit
of the Standard Model requires to have very accurate input data in order to
make a constraint for the masses of undiscovered particles, such as the Higgs
boson or other particles. In addition we can study possible deviations from
the Standard Model through anomalous couplings, CP violation, or extra
dimensions.

\section{The cross section at LO, NLO, and NNLO}
Due to the large top quark width ($\Gamma_t \approx 1.4\GeV$), the top-antitop
pair cannot hadronize into toponium resonances. The cross section therefore
appears to have a smooth line-shape showing only a moderate $1S$ peak. In
addition the top quark width serves as an infrared cutoff and as a natural
smearing over the energy. As a result, the nonperturbative QCD effects induced
by the gluon condensate are small \cite{FadinYakovlev}, allowing us to
calculate the cross section with high accuracy by using perturbative QCD even
in the threshold region. Many theoretical studies at LO and NLO have been done
in the past, and recently for the NNLO as well. The results of the NNLO
analysis are summarized in a review article~\cite{Review}. To summarize the
results for a standard approach using the pole mass, the NNLO corrections are
uncomfortably large, spoiling the possibility for the top quark mass extraction
at NLC with good accuracy because the $1S$ peak is shifted by about $0.5\GeV$
by the NNLO, the last known correction. One of the main reasons for this is
the usage of the pole mass in the calculations. It was realized that such type
of instability  is caused by the fact that the pole mass is a badly defined
object within full QCD. For more details we also refer the reader to
Ref.~\cite{YakovlevGroote}.

Let us consider the cross section of the process $e^+e^-\to t\bar t$ in the
near threshold region where the velocity $v$ of the top quark is small. It is
well-known that the conventional perturbative expansion does not work in the
non-relativistic region because of the presence of the Coulomb singularities
at small velocities $v\to 0$. The terms proportional to $(\alpha_s/v)^n$
appear due to the instantaneous Coulomb interaction between the top and the
antitop quark. The standard technique for re-summing the terms $(\alpha_s/v)^n$
consists in using the Schr\"odinger equation for the Coulomb potential to find
the Green function. The total cross section can then be related to the Green
function by using the optical theorem,
\begin{equation}\label{cross}
R=\frac{\sigma(e^+e^-\to t\bar t)}{\sigma(e^+e^-\to\mu^+\mu^-)}
 =e^2_Q\frac{72\pi}sC(r_0){\rm Im}\left[\left(1-\frac{\vec p\,^2}{3m^2}
 \right)G(r_0,r_0|E+i\Gamma)\right]\Bigg|_{r_0\to 0}
\end{equation}
where the 
Green function $G(\vec r,\vec r\,'|E+i\Gamma)$
satisfies the Schr\"odin\-ger equation
\begin{equation}
(H-E-i\Gamma)G(\vec r,\vec r\,'|E+i\Gamma)=\delta(\vec r-\vec r\,').
\end{equation}
An obstacle for a straightforward calculation at NNLO are the UV divergences
coming from relativistic corrections to the Coulomb potential (the so-called
Breit-Fermi potential). This problem can be solved by a proper factorization
of the amplitudes and by employing effective theories (see for the
review \cite{Brambilla}). The coefficient
$C(r_0)$ can be fixed by using a direct QCD calculation of the vector vertex
at NNLO in the so-called intermediate
region~\cite{Czarnecki:1998vz,Beneke:1998zp} and by using the direct matching
procedure suggested in Ref.~\cite{Hoang}. In addition, the non-factorizable
corrections cancel in the total cross section but modify 
the differential one \cite{MeYaNF}.
For the numerical solution of the final equation we used the program derived
in Ref.~\cite{Yakovlev:1999ke} by one of the authors. 

\section{On the mass definitions}
The top quark mass is an input parameter of the Standard Model. Although it is
widely accepted that the quark masses are generated due to the Higgs mechanism,
the value of the mass cannot be calculated from the Standard Model. Instead,
quark masses have to be determined from the comparison of theoretical
predictions and experimental data. There is no unique definition of the quark
mass. Because the quark cannot be observed as a free particle like the
electron, the quark mass is a purely theoretical notion and depends on the
concept adopted for its definition. The best known definitions are the pole
mass and the $\overline {\rm MS}$ mass. However, both definitions are not
adequate for  the analysis of top quark production near threshold. The pole
mass should not be used because it has the renormalon ambiguity and cannot be
determined more accurately than $300-400\MeV$. The $\overline{\rm MS}$ mass is
an Euclidean mass, defined at high virtuality, and therefore destroys the
non-relativistic expansion. Instead, it was recently suggested to use
threshold masses. There are many such definitions: the low scale (LS)
mass~\cite{Bigi}, the potential subtracted (PS) mass~\cite{Beneke}, and
one half of the perturbative mass of a fictitious $1^3S_1$ ground state
(called $1S$ mass)~\cite{HoangTeubner}. All of them could be used in
application but have to be related to the fixed notation, e.g.\ the
$\overline{\rm MS}$ mass. We suggest a new definition which we think is more
physically motivated. The objectives in defining such a mass are that
the mass should be ``short distance'', being free from soft QCD effects and
adequate  for the threshold calculations. Furthermore, the definition should
be gauge independent and well-defined within quantum field theory so that
radiative and relativistic corrections can be calculated in a systematic way.
Our definition given in~\cite{YakovlevGroote} is
$m_{\overline{\rm PS}}=m_{\rm pole}-\delta m_{\overline{\rm PS}}$ with
$\delta m_{\overline{\rm PS}}=\Sigma_{\rm soft}(\slp)|_{\slp=m_{\rm pole}}$
where $\Sigma_{\rm soft}$ is the soft part of the heavy quark self energy
which is defined as the part where at least one of the heavy quark propagators
is on-shell. Summarizing all contribution up to NNLO accuracy, we obtain for
the difference
$m_{\overline{\rm PS}}(\mu_f)-m_{\rm pole}$~\cite{YakovlevGroote}
\begin{eqnarray}\label{psmass}
-\frac{\alpha_s(\mu)C_F}\pi\mu_f\Bigg\{1-\frac{\mu_f^2}{2m^2}
  +\frac{\alpha_s(\mu)}{4\pi}\left(C_1+C_A\frac{\pi^2}2\frac{\mu_f}m\right)
  +C_2\pfrac{\alpha_s(\mu)}{4\pi}^2\Bigg\} 
\end{eqnarray}
where the coefficients $C_1$ and $C_2$ are given in Ref.~\cite{Beneke},
$m$ is the pole mass, $\mu$ is the renormalization scale, and $\mu_f$ is the
factorization scale. We have calculated the coefficients to higher orders in
$1/m$ which are explicitly given in Eq.~(\ref{psmass}) and which are recoil
corrections to the previous one. Remark that there is no contribution of the
leading order to $1/m$. Our result can be represented in a condensed form as
$m_{\overline{\rm PS}}(\mu_f)-m_{\rm pole}=-\frac12\int^{\mu_f}
\left(V_C(|\vec k|)+V_R(|\vec k|)+V_{NA}(|\vec k|)\right)d^3k/(2\pi)^3$ where
the first term $V_C$ is the static Coulomb potential, $V_R$ is the relativistic
correction (which is related to the Breit-Fermi potential but does not coincide
with it), and $V_{NA}$ is the non-abelian correction. Using three-loop
relations between the pole mass and the $\overline{\rm MS}$ mass given in
Ref.~\cite{MelRit,Steinhauser}, we fix the $\overline{\rm MS}$ mass to
take the values $\overline{m}(\overline{m})=160\GeV$, $165\GeV$, and $170\GeV$
and determine the pole mass at LO, NLO, and NNLO. This pole mass is then used
as input parameter $m$ in Eq.~(\ref{psmass}) to determine the PS and
$\overline{\rm PS}$ masses at LO, NLO, and NNLO. The obtained values for the
PS and $\overline{\rm PS}$ mass differ only in NNLO.
\begin{figure}
\psfig{figure = 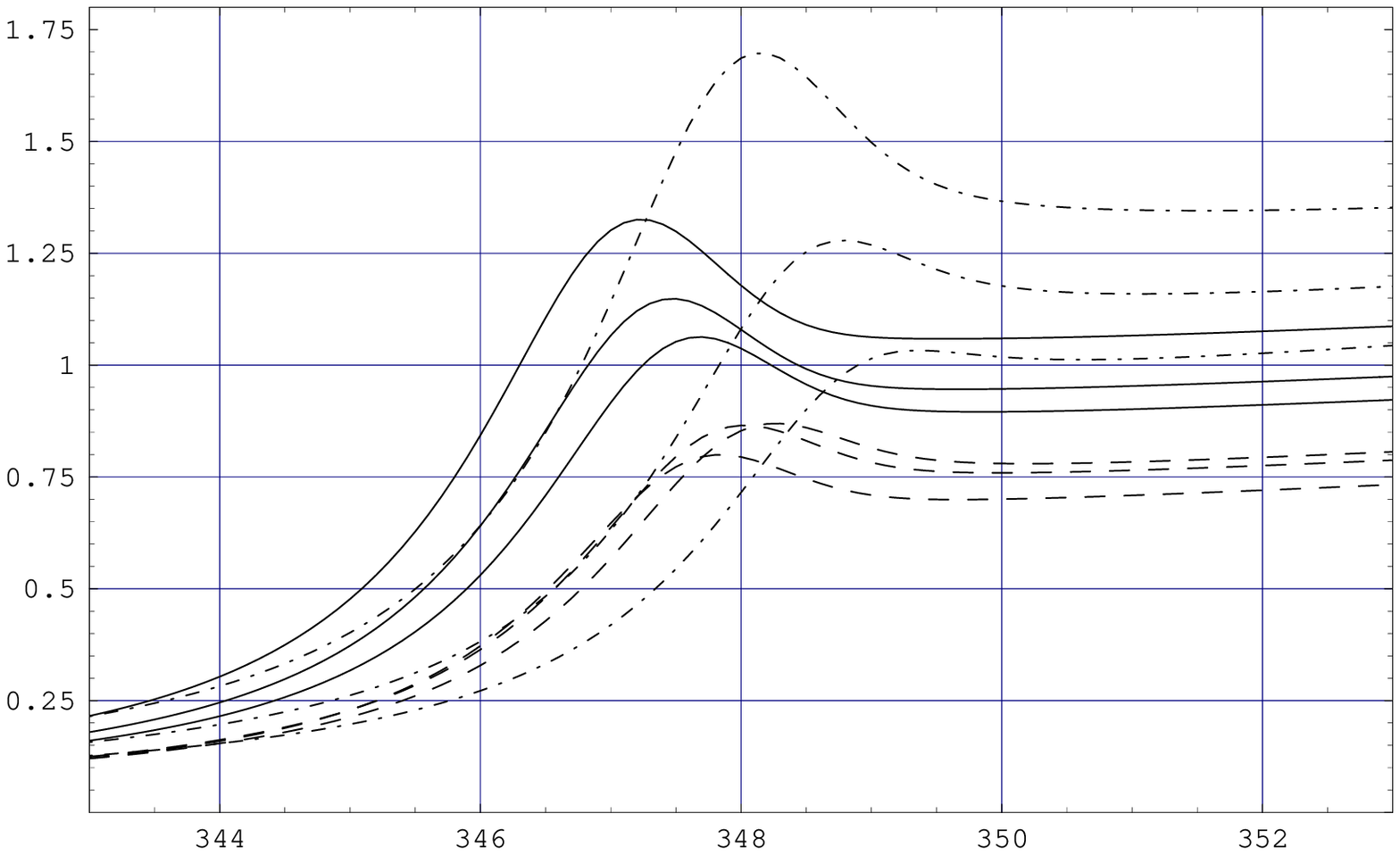,height=1.6in}
\hspace{7mm}
\psfig{figure = 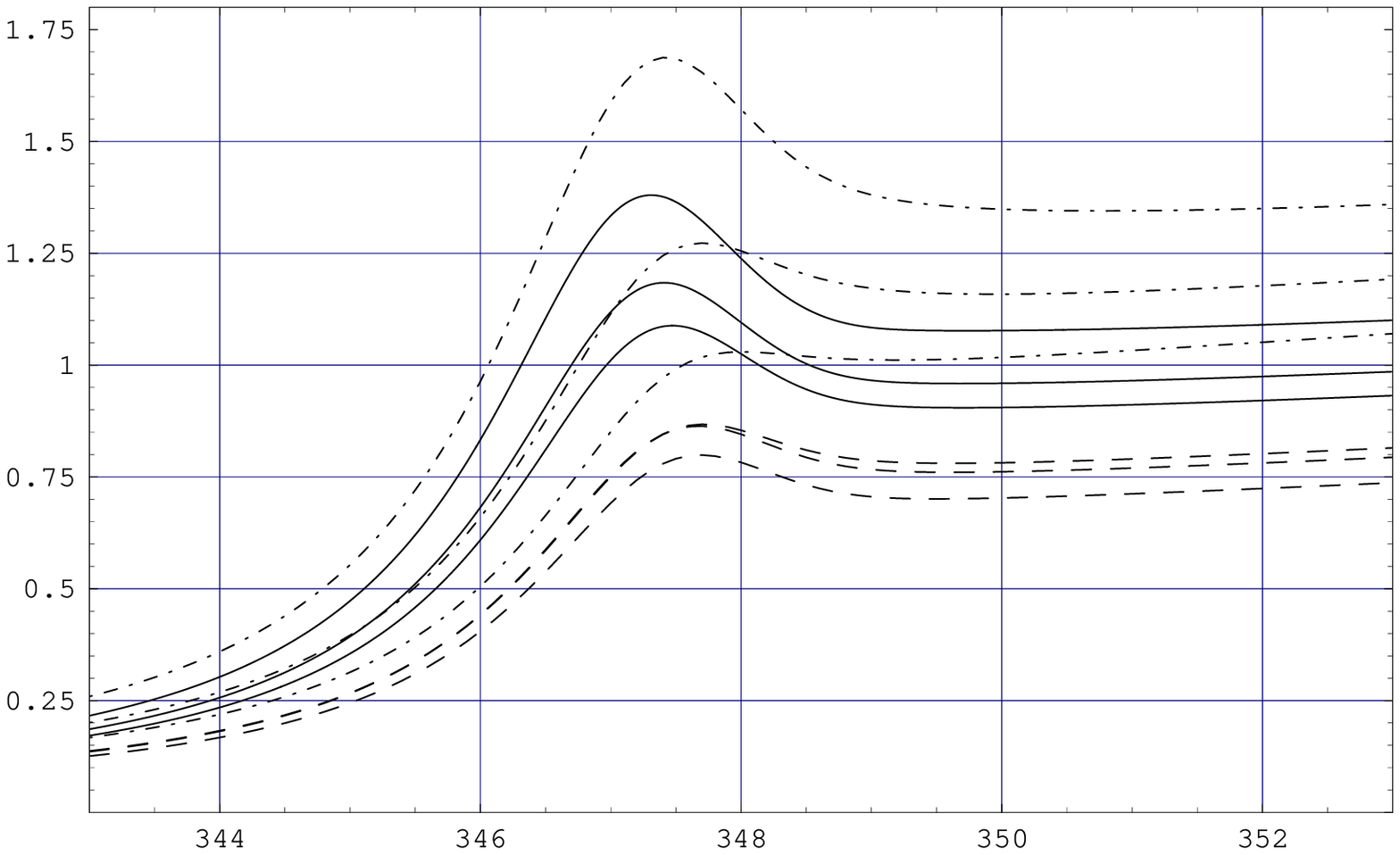,height=1.6in}
\caption{\label{fig1}NNLO threshold analysis for pole mass (left) and
  $\overline{\rm PS}$ mass (right)}
\end{figure}
Fig.~\ref{fig1} shows the results of the analysis of the cross section. The
triple of curves indicates different values for the renormalization scale,
shown for LO (dashed-dotted), NLO (dashed), and NNLO (solid). It is obvious
that in using the $\overline{\rm PS}$ mass (as well as the other threshold
masses) we gain a remarkable improvement of the stability in going from LO to
NLO to NNLO (shifts are below $0.1\GeV$). This understanding removes one of the
obstacles for an accurate top mass measurement and one can expect that the top
quark mass will be extracted from a threshold scan at NLC with an accuracy of
about $100\MeV$.

{\bf Acknowledgements:} O.Y.\ acknowledges support from the US
Department of Energy (DOE). S.G.\ acknowledges a grant given by the DFG,
Germany.

\end{document}